\documentstyle[sprocl,psfig,epsf]{article} 

\def\Journal#1#2#3#4{{#1} {\bf #2}, #3 (#4)}

\def\NPA{{\em Nucl. Phys.} A}
\def\NPB{{\em Nucl. Phys.} B}

\def\PLB{{\em Phys. Lett.} B}

\def\PRL{\em Phys. Rev. Lett.}

\def\PRD{{\em Phys. Rev.} D}
\def\PRC{{\em Phys. Rev.} C}

\def\be{\begin{equation}}
\def\ee{\end{equation}}
\def\bea{\begin{eqnarray}}
\def\eea{\end{eqnarray}}
\newcommand{\beq}{\begin{equation}}
\newcommand{\eeq}{\end{equation}}
\newcommand{\beqa}{\begin{eqnarray}}
\newcommand{\eeqa}{\end{eqnarray}}

\begin{document}

\hfill {{\small FZJ-IKP(TH)-1998-25}} 

\smallskip   

\title{EFFECTIVE FIELD THEORY APPROACHES TO PION PRODUCTION IN 
PROTON--PROTON 
COLLISIONS\footnote{Plenary talk at BARYONS~98,
Bonn, Germany, September 1998.}
} 

\author{Ulf-G. Mei{\ss}ner}

\address{FZ J\"ulich, IKP (Th), 
D-52425 J\"ulich, Germany\\E-mail: Ulf-G.Meissner@fz-juelich.de}

\maketitle\abstracts{I critically review the status of computations of 
threshold pion production in proton--proton
collisions in the framework of effective field theory approaches or
variants thereof. I also present the results of a novel diagrammatic
scheme.
}


\section{The problem}
Over the last years, very precise data on pion, $\eta$ and $\eta '$
meson production in proton-proton collisions in the threshold region
have been obtained at IUCF, CELSIUS and COSY (for a review, see 
the talk by H.O. Meyer~\cite{HOM}).
The basic process is  $pp \to pNM$, where $N$ denotes the
nucleon and $M$ a pseudoscalar meson, in our case the $\pi^0$,
$\pi^+$, $\eta$ or the $\eta '$. At the respective threshold, the produced
meson is soft, i.e. has vanishing three momentum. Consequently, in the case
of the pions, which are believed to be the Goldstone bosons related to the
spontaneous chiral symmetry breaking QCD is assumed to undergo, this process
appears to be a good testing ground for chiral perturbation theory methods.
To be specific, at threshold one has only S--waves and thus the pertinent
T--matrix for the specific case of $\pi^0$ production is parametrized in terms
of one single amplitude,
\beq\label{T}
{\rm T}^{\rm cm}_{\rm th} (pp\to pp\pi^0) = {\cal A} \, (i\,
\vec{\sigma}_1 - i \,\vec{\sigma}_2 + \vec{\sigma}_1 \times
\vec{\sigma}_2 ) \cdot \vec{p} \quad .
\eeq
The $\vec{\sigma}_{1,2}$ are the spin--matrices of the two protons. 
The amplitude is a pseudoscalar  quantity and due to  the Pauli principle,
we  are  dealing with a $^3P_0~\to {^1S}_0~s$ transition (where the '$s$'
refers to the pion angular momentum). The value
of the proton cm momentum to produce a neutral pion at rest is given by
\beq |\vec{p}\,| = \sqrt{M_\pi (m + M_\pi/4)}
\simeq \sqrt{M_\pi m} = 362.2~{\rm MeV}~,
\eeq
with $m = 938.27$~MeV the proton and $M_\pi = 134.97$~MeV the neutral pion
mass, respectively. Obviously, $|\vec{p}\,|$ vanishes in the chiral limit of
zero pion mass. Therefore the soft--pion theorem which requires a vanishing
threshold T--matrix in the chiral limit $M_\pi=0$ is trivially fulfilled (as
long as ${\cal A}$ does not become singular). Stated differently, there is
no low--energy theorem for the threshold amplitude ${\cal A}$.
Note also that the value of the
threshold momentum is already relatively large. Furthermore, it is known that
in the threshold region the strong final--state interactions
(FSI) govern the energy dependence.\cite{MuS} 
In fact, in the approaches I will discuss
in what follows, the complete amplitude is written as
\beq
T = T^{\rm ISI} \cdot T^{\rm Prod} \cdot T^{\rm FSI}~,
\eeq
where ISI denotes the initial--state interaction and the microscopic approaches
are applied to the production amplitude $T^{\rm Prod}$. Clearly, such a separation
induces a priori some model dependence, as I will discuss later. Let me first
describe the existing EFT approaches to calculate/constrain the production 
amplitude.

\section{Heavy fermion techniques at work}
Most work so far has focused on the production of neutral pions simply
because the production amplitude involves, besides many other processes,
isoscalar pion--nucleon scattering (see fig.1b), which is known to be suppressed due to
chiral symmetry. However, it should be made clear from the start that there
must also be important short distance physics, see e.g.~\cite{LR} and as 
depicted in fig.1c.
\begin{figure}[htb]
\hspace{2cm}
\psfig{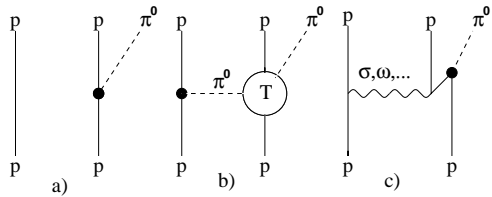}
\caption{Different mechanism contributing to the meson production operator.
a), b) and c) are referred to as the direct, rescattering and 
heavy meson--exchange terms, in order. 
\label{fig:graphs}}
\end{figure}
\noindent The challenge
of working out the pion exchange terms in chiral perturbation theory has been
been taken up by two groups, one at Seattle~\cite{bira} 
and the other at South Carolina.\cite{fred}
Their calculational schemes are similar. To be definite, I concentrate in what follows
on the work of the Seattle group. They use the effective pion--nucleon
Lagrangian to second order in small momenta (i.e. tree level graphs), 
the appearing low--energy constants (LECs)
have already been determined from pion--nucleon scattering data.\cite{bkmlec}
The nucleons are treated as very heavy spin--1/2 fields with  four--momentum
\beq
p_\mu = m v_\mu + k_\mu~,
\eeq
with $v_\mu$ the four--velocity subject to the constraint $v^2=1$ and $v\cdot k
\ll m$. To leading order, the nucleon propagator is $S(k) = i / (v\cdot k)$ and one
can set up a consistent power counting scheme in terms of small momenta, the pion
mass and inverse powers of the nucleon mass.
However, their calculations involve a modified counting scheme due to the abovementioned
fact that the momentum does not scale like the pion mass, so the heavy fermion
propagator for a small residual momentum $k_\mu$ includes the first kinetic energy
correction
\beq\label{prop}
\frac{i}{v\cdot k}\to \frac{i}{k_0 + \vec{k}^2/2m}~,
\eeq
since in the rest frame $v_\mu = (1, \vec{0}\,)$.
The necessity of such a modification follows from the fact that
the heavy baryon formalism can not cope with
external momenta as large as $|\vec p\,| \simeq \sqrt{ m M_\pi}$. 
Let me demonstrate the problems of the heavy
baryon approach for a simple (but generic) example. Consider a Feynman diagram
which involves a propagating nucleon after emission of the real $\pi^0$. I
show that this nucleon propagator can not be expanded in powers of $1/m$ in the
usual way. Let $v^\mu=(1,\vec{0} \,)$ be the four--vector which selects the
center--of--mass frame. The four--vector of the propagating nucleon is
$m \, v + k$ with $k^\mu=
(-M_\pi/2, \vec p\,)$  and $k^2 = - m M_\pi$. Starting on the left hand side
with the correct result based on a fully relativistic calculation
and then performing the usual $1/m$
expansion of the heavy baryon formalism, one has
\begin{equation}
-{1\over M_\pi} = {1\over
v\cdot k +k^2/2m} = {1\over v\cdot k} \sum_{n=0}^\infty \bigg( {- k^2 \over 2m
\, v\cdot k} \bigg)^n = - {2 \over M_\pi} \sum_{n=0}^\infty (-1)^n~.
\end{equation}
One sees that infinitely many terms of the  $1/m$--expansion contribute to
the same order. The resulting series does not even converge and oscillates
between zero and twice the correct answer. The source of this problem is the
extreme kinematics of the reaction $NN\to NN \pi$ with $|\vec p\,| \simeq
\sqrt{m M_\pi}$. In that case the leading order operator ${\cal O}^{(1)} = i\,
v \cdot \partial $ and the next--to--leading order operator ${\cal O}^{(2)} =
- \partial \cdot \partial/2m$ lead to the same result, here, $\pm M_\pi/2$.
This problem is merely related to ``trivial'' kinematics. Therefore, one
either has to modify the propagator as described above or calculate fully
relativistically, as discussed in the next section. Space forbids to discuss
in big detail the results of these studies, I rather concentrate on the most
intriguing ones. First, one finds that the rescattering contribution (fig.1b)
comes out with an opposite sign to what one gets in meson--exchange models,
i.e. it interferes
constructively with the direct production in the J\"ulich 
meson--exchange model~\cite{unsers} and 
destructively in the chiral framework, respectively. Note that while there is
still debate about the actual numerical treatment 
and the ensuing size of the rescattering contribution
in the chiral perturbation theory approaches,\cite{lee} the sign
difference with the meson--exchange model can be considered a genuine
feature. This point was addressed in ref.\cite{J2}.
It was argued that the treatment underlying the isoscalar
pion--nucleon scattering amplitude and the related transition operator
for the process $NN \to NN\pi$ in the chiral framework is not yet 
sufficiently accurate and thus the resulting rescattering contribution
should be considered an artifact of this approximation. Clearly, this 
does not mean that chiral perturbation theory is invalid but rather
that higher order  (one loop) effects need to be accounted for. 
Second, to account for the data, one has to include short distance
physics, like e.g. heavy meson exchanges (see fig.1c) and also 
meson--exchange currents, in particular the anomalous $\pi \rho \omega$
vertex. A typical result is shown in fig.2, which also shows that there
is still a sizeable uncertainty related to e.g. the treatment of the 
isoscalar rescattering. 
\begin{figure}[htb]
\hspace{1.8cm}
\psfig{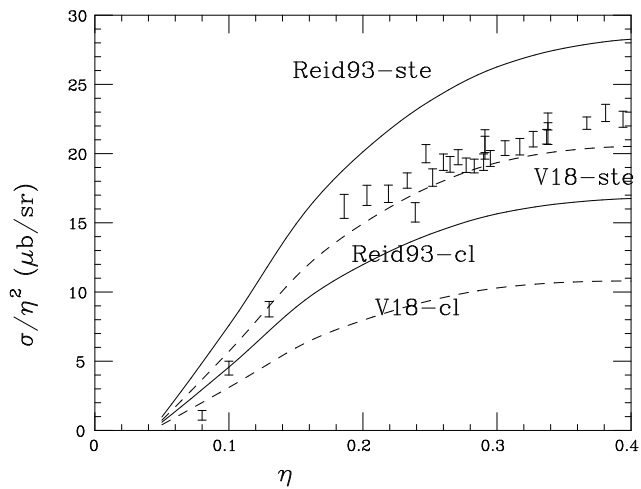}
\caption{Cross--section for $pp\to pp\pi^0$ as function of the pion momentum
$\eta$ in units of $M_\pi$ for two NN potentials (Argonne V18 and Reid93) and two
parameter sets for the isoscalar $\pi N$ amplitude (ste and cl). Figure
courtesy of Bira van Kolck.
\label{fig:bira}}
\end{figure}

The first one--loop calculation (to third order in small momenta) was perfomed
by Gedalin et al.\cite{isra}.  They work in the conventional heavy baryon 
framework~\cite{bkmrev} using the lowest order fermion propagator (see the
left--hand side of eq.(\ref{prop})). Some typical one--loop graphs are
shown in fig.3. There are, of course, many more diagrams, most of them giving
rise to mass and coupling constant renormalization. At this order, there appears
\begin{figure}[htb]
\hspace{2.2cm}
\psfig{figure=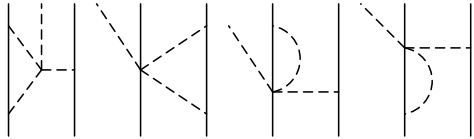,height=.7in}
\caption{Class of one loop graphs involving the $\pi\pi$ interaction. Dashed/solid
lines denote pions/nucleons.
\label{fig:loops}}
\end{figure}
also a four--nucleon--pion contact term with a LEC, called $d_1$. This LEC
could in principle be fitted from one total cross section point. 
The authors of ref.\cite{isra}
estimate its values from resonance saturation, specifically by $\rho$
and $\omega$ exchange. 
This procedure induces, of course,
some model--dependence. (The status of resonance saturation in the one--nucleon
sector is discussed in ref.\cite{bkmlec}) Therefore, the total cross section
can be predicted without any free parameters and it agrees quite nicely with
the data in the threshold region. However, there are a couple of loopholes
with this result. First, as discussed before, using the leading order
nucleon propagator, one can not expect a convergent series. This is reflected
in the results of ref.\cite{isra},
\beq
T^{\rm rescattering} : T^{\rm loop} : T^{\rm direct} \simeq 5:2:1~,
\eeq
according to figs.1b,3,1a, respectively. Without the loop contribution, the total
cross section is underestimated by a factor of about 2.5.
One can pin down this
problem of the bad convergence more specifically. The class of loop graphs
shown in fig.3 is directly proportional to the scalar form factor of the
nucleon at $t = M_\pi m \simeq 0.1\,$GeV$^2$. In the heavy fermion approach, it
leads to a very large enhancement of the production amplitude. If one, however,
calculates these loop diagrams fully relativistically and evaluates the scalar
form factor, its contribution is very small.\cite{bkmr} This simply reflects
the fact that a one--loop calculation at third order is not sufficiently accurate,
even the shift of the scalar form factor to the Cheng--Dashen point at $t = 4M_\pi^2
\simeq 0.02\,$GeV$^2$ is off by a factor of two. Stated differently, the relativistic
calculation shows that higher order $1/m$ terms are large.
Similar remarks apply to the
full isoscalar pion--nucleon scattering amplitude. From these results I conclude
that heavy baryon chiral perturbation theory is presumably not 
the appropriate framework
to gain a deeper understanding of the role of chiral dynamics in the process
$pp \to pp\pi^0$. As already pointed out in ref.\cite{J2}, a more detailed study
of the  reaction $pp \to d\pi^+$ could, however,
 pave the way to a deeper understanding of the pion
dominated part of the transition operator (simply because short--distance physics
plays a much smaller role in this process).
Finally, I remark that the treatment discussed so far  involves an unavoidable
ambiguity due to  the exchanged pion (say in the rescattering graph, fig.1b)
being off--shell. For a given prescription of defining the pion field, one can perform
an off--shell extrapolation but this will always be model--dependent -- in quantum
field theory  only on--shell matrix elements and transition currents can be
calculated.

\section{Diagrammatic approach}

Based on the observation that the heavy baryon framework is not expected to converge,
in ref.\cite{bkmr} a different approach was pursued. Consider first the process
$pp\to pp\pi^0$. Approximating the near threshold T--matrix by the T--matrix exactly at
threshold one gets for the unpolarized total cross section
\begin{equation}\label{stots}
\sigma_{\rm tot}(T_{\rm lab}) = |{\cal A}|^2 \, \int dW \, KF(W) \, F_p(W)~,
\end{equation}
where the flux and three-body phase space factors, denoted as $KF (W)$,
can be approximated by an analytical expression which is
accurate within a few percent in the threshold region. 
$F_p(W)$ is the correction factor due the final--state interaction,
with $W$ the final--state  invariant di--proton mass. It
is evaluated in the  effective range approximation,
\beq
F_p (W) = \biggl\{ 1 + \frac{a_p}{4}(a_p+r_p)P^2_* + \frac{a_p^2 r_p^2}{64}
P_*^4 \biggr\}^{-1}~,
\eeq
with $P_*^2 =W^2 - 4m^2$ and
$a_p$ ($r_p)$ the $pp$ scattering length (effective  range) including  electromagnetic
corrections.  This is of course a very
strong assumption but it allows one to explain the energy dependence of the
experimental total cross sections very accurately in terms of a single constant
amplitude ${\cal A}$. At the end of this section, I derive this particular
treatment of the FSI from an effective  field theory
(EFT) approach. Separating off the final--state interaction
in that way, one can then pursue a diagrammatic approach to the (on-shell)
production amplitude ${\cal A}$. This  allows one  to investigate in a simple
fashion the role of one-pion exchange and chiral loop effects together with
shorter range exchanges due to heavier mesons.  In a similar fashion, one
can investigate the other processes $pp\to pn\pi^+$, $pp \to pp\eta$ and
$pp \to pp\eta '$ (see below).
To be specific,
consider  the $pp\pi^0$ reaction. Assuming the form eq.(\ref{stots}),
the S--wave threshold amplitude can be extracted from the data,
\beq\label{Aexp}
{\cal A}^{(\rm emp)} = (2.7- i\, 0.3)~{\rm fm}^4~,
\eeq
as shown by the solid line in fig.4. The imaginary part is due to the
\begin{figure}[htb]
\hspace{1.6cm}
\psfig{figure=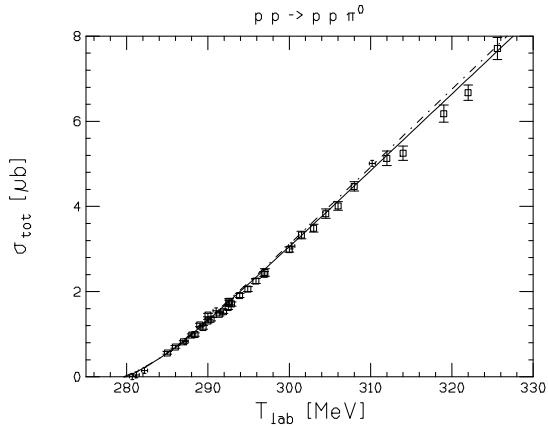,height=2.2in}
\caption{
Fit to the total cross section for $pp \to pp\pi^0$ as described in the
text (solid line). The data are from IUCF (boxes) and
CELSIUS (crosses). The dashed line is result of the diagrammatic approach.
\label{fig:pppi0}}
\end{figure}
$^3P_0$ $pp$ phase shift  taken at the threshold energy in the
lab frame, $T^{\rm lab}_{\rm th} = 279.65$~MeV, where $\delta(^3P_0) =
-6.3^\circ$. Thus the imaginary part Im\,${\cal A}$ is about $-1/9$ of the
real part Re\,${\cal A}$ and contributes negligibly to the total cross section
near threshold. This number can be well understood in terms of chiral $\pi^0$
exchange (including chiral $\pi^0$ rescattering) and heavy meson ($\omega,\,
\rho^0,\, \eta$) exchanges based on a relativistic Feynman diagram
calculation with all parameters being fixed from reliable methods, like
forward NN dispersion relations for the $\omega$ and a dispersion--theoretical
analysis of $\bar{N}N \to \pi\pi$ for the $\rho$. Interestingly, in the
relativistic approach the rescattering contribution interferes constructively
with the direct production term (cf. fig.1a), in contrast with  the heavy baryon
approach and in agreement with the meson--exchange models.
One can also evaluate some classes of one-loop graphs and finds that
they lead to small corrections of the order of a few percent.
Therefore chiral loops do not seem to play any significant role
in the processes $NN\to NN \pi$,  which are dominated by one--pion exchange
and short--range physics. I remark also that
both the long range $\pi^0$ exchange and the short range vector
meson exchange lead to contributions to the threshold amplitude ${\cal A}$
which do not vanish in the chiral limit $M_\pi \to 0$. There is no
chiral suppression of the reaction $pp\to pp \pi^0$ compared to other $NN\pi$
channels.  In all cases the respective threshold amplitudes are non-zero (and
finite) in the chiral limit. This is in contrast to the widespread believe
that $pp \to pp \pi^0$ is suppressed for reasons of chiral symmetry.

 Within the same approach, one can  investigate the threshold
behavior of the process $pp \to pn\pi^+$. It is given in terms of ${\cal A}$
and the triplet threshold amplitude ${\cal B}$,
\beq\label{T+}
{\rm T}^{\rm cm}_{\rm th} (pp\to pn\pi^+) = \frac{{\cal A}}{\sqrt{2}} \, (i\,
\vec{\sigma}_1 - i \,\vec{\sigma}_2 + \vec{\sigma}_1 \times
\vec{\sigma}_2 ) \cdot \vec{p} - \sqrt{2}{\cal B} i (\vec{\sigma}_1 + 
\vec{\sigma}_2 ) \cdot \vec{p}~.
\eeq
Here, the second  term refers to the  $^3P_1 \to ^3S_1 s$  transition. Note also
that the singlet transition is suppressed because of  the large singlet
scattering length.
From the IUCF data, one can determine the empirical value of the triplet
amplitude, ${\cal B}^{\rm emp} = (2.8 -i1.5)\,$fm$^4$. The large imaginary
part is related to the strong ISI in the $^3P_1$ entrance channel, which naturally
can not be explained by tree graphs only. The value of the $^3P_1$ phase at
the threshold energy for $pp\to pn\pi^+$ is $-28.1^\circ$.
The corresponding real part Re~${\cal B}$
is well reproduced by chiral one--pion exchange and short--range vector meson
physics. For the same parameters as used in the study of $pp\pi^0$, one gets
Re~${\cal B} = 2.74\,$fm$^4$. A more detailed discussion of  this channel and  the
problems related to the large imaginary  part can be found in ref.\cite{bkmr}.

Let me now turn to $\eta$ production. The threshold matrix element takes the
form
\begin{equation}
{\rm T}^{\rm cm}_{\rm th}(pp\to pp \eta) = {\cal C} \, ( i\,\vec
\sigma_1 - i\, \vec \sigma_2+\vec \sigma_1 \times \vec \sigma_2)\cdot \vec p
\,. \end{equation}
with ${\cal C}$ the (complex) threshold amplitude for $\eta$--production. The
$\eta$--production threshold is reached at a proton laboratory kinetic energy 
$T_{\rm lab}^{\rm th} = M_\eta(2+M_\eta/2m) = 1254.6$ MeV, where
$M_\eta=547.45$ MeV denotes the eta--meson mass.
In the case of $\eta$--production near threshold it is also important to
take into account the $\eta p$ final--state interaction, since the $\eta
N$--system interacts rather strongly near threshold. In fact a recent 
coupled--channel analysis~\cite{batinic} of the $(\pi N, \eta N)$-system  finds
for the real part of the $\eta N$ scattering length Re$\,a_{\eta N} =(0.717\pm 
0.030)$ fm. For comparison, this value is about six times larger than the $\pi^-p$
scattering length, $a_{\pi^- p} = 0.125$ fm, as  measured e.g. in pionic
hydrogen.  In ref.\cite{bkmr}, it is 
assumed that the correction due to the S--wave $\eta p$ FSI 
near threshold can be treated in effective range approximation
analogous to the S--wave $pp$ FSI. The further assumption is made
that the FSI in the $pp$ subsystem and in the two $\eta p$ subsystems do not 
influence each other and that they factorize. The corresponding form
of  the unpolarized total cross section in terms of the S--wave amplitude ${\cal C}$
the various FSI functions $F_p (W)$, $F_\eta (s_\eta)$ can be found in ref.\cite{bkmr}.
Note that the $\eta p$ FSI function is complex--valued (because the
$\eta N$ scattering length has a sizeable imaginary part). From the CELSIUS data, 
one finds for the modulus of  the threshold amplitude  
$|{\cal C} | = 1.32 \, {\rm fm}^4$.
\begin{figure}[htb]
\hspace{1.6cm}
\psfig{figure=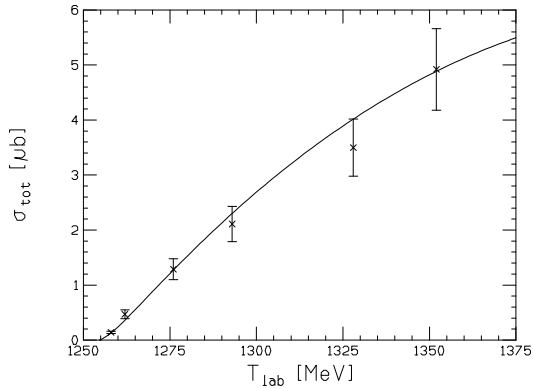,height=2.in}
\caption{
The eta--production cross section $\sigma_{\rm tot}(pp \to pp \eta)$ as a 
function of $T_{\rm lab}$. The  data  are taken from CELSIUS.
\label{fig:ppeta}}
\end{figure}
The resulting energy dependent cross section
from threshold up to $T_{\rm lab} = 1375$ MeV is shown in fig.~\ref{fig:ppeta} 
together with the data from CELSIUS.
It is rather astonishing that one can describe the total
cross section data up to 100 MeV above threshold with a constant threshold
amplitude ${\cal C}$ and a simple factorization ansatz for the three--body FSI.  
The relativistic Feynman graphs contributing to the threshold amplitude
${\cal C}$ can be easily evaluated. To account for the strong $\eta N$
S-wave rescattering, which is often attributed to the nucleon resonance
$S_{11} (1535)$, one has to introduce a local $NN\eta\eta$ contact term,
${\cal L}_{\eta N} = K \bar{N}(x) N(x) \eta^2 (x)$. Its strength $K$ can be 
obtained from fitting the scattering length $a_{\eta N}$. With $g_{\eta N} = 
5.3$, which is close to flavor SU(3) estimates and determinations from
boson--exchange models, one can exactly reproduce the value of ${\cal C}$.
It is worth  pointing out that $\rho^0$ exchange is the dominant contribution,
because it is enhanced by factors of $M_\eta / m \simeq 0.6$ quite in contrast
to the neutral pion case, where $\omega$ and $\pi$ exchange are the dominant
mechanisms.  One can also predict the threshold S--wave  amplitude ${\cal D}$
for the process $pn \to pn\eta$. One finds $|{\cal D}| = 1.8\,$fm$^4$, somewhat
below the experimental value,  $|{\cal D}|^{\rm emp} 
= 2.3\,$fm$^4$, as determined from the recent
data on  quasifree production off the deuteron  measured at CELSIUS.

At COSY, $\eta '$ production very close to threshold has been measured for the
first time, at much
smaller excess energies  than the few data points from the now defunct 
SATURNE facility.
These data seem to follow three--body phase space with little indication of
FSI. Note, however, that so close to threshold  the energy  determination is
difficult  and thus there is a sizeable uncertainty in the value of the excess
energy (for more details, see ref.\cite{mos}). Within the diagrammatic approach,
the empirical S--wave  amplitude can be obtained with 
\beq
g_{\eta ' NN} \, (1 - 1.28\varepsilon ) = 1.12~,
\eeq
with  $g_{\eta ' NN}$ the so far undetermined $\eta ' - N$ coupling  constant
and $\varepsilon$ the pseudoscalar to pseudovector mixing parameter. Since
the $\eta '$ is not a  Goldstone boson, one is not forced to use only derivative
couplings. Interestingly, only the tensor interaction of the
$\rho$--exchange ($\sim \kappa_\rho$) is sensitive to the parameter $\varepsilon$. 
Combining this result with constraints from deep inelastic 
lepton--nucleon scattering~\cite{TGV}  (the spin content of the nucleon),
one infers
\beq
g_{\eta ' NN} = 2.5 \pm 0.7~, \quad \varepsilon = 0.4 \pm 0.1~.
\eeq
It remains to be
seen whether other $\eta'$--production processes (e.g. photoproduction $\gamma
p \to \eta' p $) are consistent with these values.

Finally, I want to give an elementary derivation of the final-state
interaction correction factor  $F_p(W) = [1+a_p^2 P_*^2]^{-1}, \,P_*^2= 
W^2/4 -m^2$, in the scattering length approximation (i.e. for the
effective range $r_p=0$). Close to 
threshold all final state three--momenta are small and therefore one can
approximate both the meson production process $NN\to NN\pi^0$ and elastic 
$NN\to NN$ scattering by momentum independent contact vertices proportional to
$\cal A$ and the scattering length $a_p$, respectively. Consider first low
energy $NN$--scattering in this approximation. The bubble diagrams with
0,1,2,\dots rescatterings can be easily summed up in the form of a geometric 
series,
\begin{equation} a_p - 4 \pi a_p^2 \int{d^3l\over (2\pi)^3} {1\over
P_*^2+i0^+-l^2} + \dots = a_p + i\, a_p^2 P_* + \dots = {a_p \over 1 - i\, a_p
P_*} \,\,, \end{equation} 
using dimensional regularization (in  the $\overline{MS}$--scheme)
to evaluate the (vanishing) real part of the
loop integral. Obviously, the sum of these infinitely many loop diagrams is
just the unitarized scattering length approximation which leads to 
$\tan \delta_0(W) = a_p P_*\,$. 
Next, consider in the same approximation meson production followed by an
arbitrary number of  $NN$--rescatterings in the final--state.  Again, these
loop diagrams can be summed up to, 
\begin{equation} {{\cal A} \over 1 - i \, a_p P_*}\,\,, \end{equation}
and taking the absolute square, 
\begin{equation} \bigg|{{\cal A} \over 1 - i\, a_p P_*} \bigg|^2 = {|{\cal
A}|^2 \over 1 + a_p^2 P_*^2} \,\,, \end{equation} 
one encounters the final--state interaction correction factor $F_p(W)=[1+a_p^2
(W^2/4-m^2)]^{-1}$ in scattering length approximation. Since the scattering
length is much bigger than the effective range parameter for $NN$--scattering
($a_p >> r_p$) one has already derived the dominant effect due to the FSI.
Of course, in order to be more accurate one should eventually go
beyond momentum independent contact vertices. The main point, however,
I want to make
here is that the FSI correction factor $F_p(W)$ (for
$r_p=0$) has a sound foundation in effective field theory.   

Of course, this approach also has some drawbacks. First, in view of the
recent polarization measurements   at IUCF and RCNP, one should go beyond
the S--wave approximation. Second, it  is not obvious how to systematically
improve the calculations and third, the  treatment of the FSI might be too
simplistic (for a recent discussion, see e.g. ref.\cite{mecker}).

\section{Outlook}

I given a short review of EFT or EFT--inspired approaches to pion (as well
as $\eta$ and $\eta '$) production in proton--proton collisions. The much discussed
reaction $pp \to pp\pi^0$ does not seem to offer a testing  ground for chiral
dynamics due to large  uncertanties related to chiral pion--exchange, the
short distance physics and the strong  ISI/FSI. Even the novel EFT scheme due
to Kaplan, Savage  and Wise~\cite{david} is (in its present formulation)
not able to cope with pion momenta as large  as in  this reaction. On the
other hand, there exist also the rather successfull meson--exchange
models based on effective Lagrangians from the J\"ulich (and RCNP) group(s).
While  these work fairly  well, there is no systematics means of improving them
and also, they employ  highly  model--dependent meson--nucleon form  factors,
which have no sound physical basis. Since  a consistent EFT approach to deal
with these processes does not seem to be on the horizon, it might  be worthwhile
to try to constrain the meson--exchange models with EFT   ideas and eventually
gain a better understanding  of the meson--nucleon interaction regions.
Also, more effort should be invested in studying the process $pp\to d\pi^+$
since it  is much less contaminated by short--distance physics. Theory still
has a long  way to go to catch up with the tremendous  precision obtained
in the recent  experiments.  

\section*{Acknowledgments}
I thank Dan-Olof Riska for giving me the opportunity to present these thoughts
and Silas Beane for a critical reading of the manuscript. I also thank the Institute
for Nuclear Theory at the University of Washington for its hospitality and the
Department of Energy for partial support during the completion of this work.

\section*{References} 

\end{document}